\documentclass[preprint,prb]{revtex4}

\usepackage{graphicx}   
\usepackage{bm}
\usepackage{epsfig}
\usepackage{subfigure}  
\usepackage{amsmath}    
\usepackage{verbatim}   
\usepackage{hyperref}   
\usepackage{color}      
\usepackage{bbding}
\usepackage{pifont}
\usepackage{amsfonts}
\usepackage{pdfsync}

\newcommand{\bfr}{ {\bf r}} 
\newcommand{\bfrp}{ {\bf r'}} 
\newcommand{\bfrpp}{ {\bf r''}} 

\newcommand{\vecr}{{\mathbf{r}}}

\newcommand{\abs}[1]{|#1|} 
\newcommand{\brokt}[3]{\left\langle #1 \right| #2 \left| #3\right\rangle}

\begin{document}

\title{Length Dependence of Ionization Potentials of Trans-Acetylenes: Internally-Consistent DFT/GW Approach}
\author{Max Pinheiro Jr, Marilia J. Caldas}
\affiliation{Instituto de F\'{\i}sica, Universidade de S\~ao Paulo, Caixa Postal 66318, CEP 05315-970 S\~ao Paulo SP, Brazil}
\author{Patrick Rinke}
\affiliation{COMP Centre of Excellence and Helsinki Institute of Physics, Department of Applied Physics, Aalto University,
P.O. Box 11100, FI-00076 Aalto, Espoo, Finland}
\affiliation{Fritz-Haber-Institut der Max-Planck-Gesellschaft, Berlin, D-14195 Germany}
\author{Volker Blum}
\affiliation{Department of Mechanical Engineering and Materials Science and Center for Materials Genomics, Duke University, Durham, North
Carolina 27708, United States}
\affiliation{Fritz-Haber-Institut der Max-Planck-Gesellschaft, Berlin, D-14195 Germany}
\author{Matthias Scheffler}
\affiliation{Fritz-Haber-Institut der Max-Planck-Gesellschaft, Berlin, D-14195 Germany}
\date{\today}

\begin{abstract}

We follow the evolution of the Ionization Potential (IP) for the paradigmatic quasi-one-dimensional trans-acetylene family of conjugated molecules, from short to long oligomers and to the infinite polymer trans-poly-acetylene (TPA). Our results for short oligomers are very close to experimental available data. We find that the IP varies with oligomer length and converges to the given value for TPA with a smooth, coupled inverse-length-exponential behavior. Our prediction is based on an ``internally-consistent'' scheme to adjust the exchange mixing parameter $\alpha$ of the PBEh hybrid density functional, so as to obtain a description of the electronic structure consistent with the quasiparticle approximation for the IP. This is achieved by demanding that the corresponding quasiparticle correction, in the $GW$@PBEh approximation, vanishes for the IP when evaluated at PBEh($\alpha^{ic}$). We find that $\alpha^{ic}$ is also system-dependent and converges with increasing oligomer length, allowing to capture the dependence of IP and other electronic properties.

\end{abstract}

\maketitle

\section{Introduction}

Ionization potentials (IP) and electron affinities (EA) are fundamental electronic properties of composite or complex systems and, notably in the past years when organic materials (molecular or polymeric) are being sought for applications in optoelectronic devices,\cite{cahe-kahn03am,kulk+04cm} much attention has been paid to this subject. The input from theoretical calculations is extremely relevant not only to help in gauging experimental data, but also in order to identify new directions for the optimal composition of actors in the building of a device. There has been thus a search for theoretical methods that can give us accuracy together with feasibility of calculations, spanning a wide range of both inorganic and organic molecular systems.\cite{zhan+03jpca,zhao+06jctc,zhan-musg07jpca,dabo+13pccp}
In particular, the class of linear or quasi-linear molecular systems --oligomers or polymers-- offer a special work space that allows one to concentrate on length-dependence (just one relevant dimension) of the properties of interest: indeed, the dependence and evolution of the IP, EA and electronic gaps for with structural characteristics or compositions  is a topic of intense studies.~\cite{Rade+01JMS,ruin+02prl,Hutc+03PRB,Chi-Wegn05mmrc,Zade+11acr,Li+12PSSB} For short oligomers in fixed geometries, as for small molecules, these properties can be obtained with high accuracy from  high-level quantum-chemistry calculations that go beyond the mean-field approximation, serving as benchmarks for other computational electronic structure approaches.~\cite{Luza03JSC,Musi-Bart04CPL} As the oligomer length increases, however, the computational cost of such calculations quickly becomes prohibitive.~\cite{rohl+01sm}

For such polyatomic systems, in particular for large molecules or extended materials, density-functional theory (DFT) has become the method of choice for a theoretical description, analysis, or prediction of ground state electronic properties, stable or metastable atomic structures, vibrations, and structure--property relationships.~\cite{capa-cald03prb,ferr+03prl,Li+12PSSB} We recall that, despite the fact that DFT is a ground-state theory, certain excitations that can be expressed as differences of ground-state total energies are accessible. The IP and the EA are defined as:

\begin{align}
\label{eq:IP}
IP & = {E^{N-1}} - {E^{N}} \\
\label{eq:EA}
EA & = {E^{N}} - {E^{N+1}} \ ,
\end{align}

\noindent where $E^N$, $E^{N-1}$ and $E^{N+1}$ are the total energies of the \textit{N}-, \textit{(N-1)}- and the \textit{(N+1)}-particle systems in the ground state. If $E^N$ and $E^{N\pm1}$ are computed for the same molecular geometry we obtain ``vertical excitations''. The difference

\begin{equation}
\label{eq:Egap}
{E_{gap}} = IP-EA
\end{equation}

\noindent is the electronic gap of the system, also called the self-consistent or $\Delta$SCF gap. Experimentally it is determined by direct and inverse photoemission, and should not be taken as the optical gap.

In exact DFT the values of IP and EA from Eqs.\ref{eq:IP} and \ref{eq:EA} are also given~\cite{levy+84pra,ambl-bart85prb} by the highest occupied Kohn-Sham (KS) levels of the \textit{N} and \textit{N+1} electron systems, respectively. For approximate DFT, the Slater-Janak transition states, i.e. the highest occupied KS levels of the \textit{N-1/2} and of the \textit{N+1/2} electron systems, should provide an accurate estimate of the IP and EA energies:

\begin{equation}
\label{eq:SJ_IP}
IP \approx \epsilon^{N-1/2}_N
\end{equation}

\begin{equation}
\label{eq:SJ_EA}
EA \approx \epsilon^{N+1/2}_{N+1}
\end{equation}

The difference between the highest occupied KS levels of the \textit{N}- and \textit{(N-1/2)}-  and of the \textit{(N+1/2)}- and \textit{(N+1)}-electron systems reflects the self-interaction or localization error of the highest occupied KS orbitals of the \textit{N} and \textit{(N+1)}-electron systems.~\cite{yang+12jcp} For approximate DFT the energies noted in Eqs. \ref{eq:SJ_IP} and \ref{eq:SJ_EA} should be taken.~\cite{Slat72AQC,libe00prb}

The HOMO and LUMO (highest occupied and lowest unoccupied molecular orbital) levels of  KS theory for the ground-state of a given \textit{N}-electron system, that is, the $\epsilon^{N}_N$ and $\epsilon^{N}_{N+1}$ energies, are however frequently used for the definitions of (the negative of) IP and EA, and the difference

\begin{equation}
\label{eq:HLgap}
E^{KS}_{gap} = \epsilon^{N}_{N+1} -\epsilon^{N}_N
\end{equation}

\noindent is usually termed the Kohn-Sham HOMO-LUMO gap.

As said above, the use of beyond mean-field methods for large systems is still a challenging issue, and for this reason it is not known how the IP or the electronic gap develop as a function of polymer length. While Berger {\it et al.} showed,~\cite{Berg+05JCP} through a ``dielectric needle'' model for the polymer, that the polarizability per unit mer is inversely proportional to the polymer length, no such analytic dependence is known for the ionization potential. For approximate DFT functionals the IP from Eq. \ref{eq:IP}, IP$_{\Delta SCF}$,  is usually more accurate than from the plain HOMO energy $-\epsilon^{N}_N$ because it is less affected by the self-interaction error.~\cite{Perd+82PRL} However, going from short oligomers to more extended systems, different DFT functionals give rise to different length dependence of the IP$_{\Delta SCF}$, ranging from concave and straight to convex as a function of inverse length,~\cite{Salz-Aydi11JCTC} so the problem is still under discussion.

To address this problem we will design a DFT functional consistent with many-body perturbation theory in the $GW$ approach.~\cite{hedi65pr,hedi-lund69ssp}  $GW$ has become the prime method for the computation of quasiparticle energies in solids as measured by direct or inverse photoemission~\cite{Aulb+00ssp,Onid+02rmp,Rink+08pssb} and is increasingly applied to organic systems~\cite{Blas+11PRB,Fabe+12jmsci}  including polymers.~\cite{Ethr+96PRB,rohl-loui99prl,Hors+99PRL,Hors+00PRB,rohl+01sm,Tiag+04PRB,Ferr+12PRB,Chan-Jin12JCP} The standard procedure is to apply a single iteration of the $GW$ approach ($G_0W_0$) as a many-body perturbation to the results of  a DFT or Hartree-Fock (HF) calculation. The single-particle wavefunctions of DFT or HF, the respective orbital energies and the resulting dielectric screening form the input to the $G_0W_0$ calculation and therefore determine the behavior of the screened Coulomb interaction $W_0$. The screening strength of $W$ decreases with increase of the HOMO-LUMO gap. Thus local or semi-local DFT functionals that produce a too small gap compared to the real electronic gap  would overestimate screening, whereas HF that produces a too large gap would underestimate it. As shown recently by Bruneval and Marques,~\cite{Brun-Marq13JCTC} input orbitals and energies derived from hybrid functionals with a high fraction of exact-exchange yield $G_0W_0$ IPs that agree well with experiment for small organic molecules, whereas for larger molecules the fraction of exact-exchange has to be considerably lower.~\cite{Koer-Maro12prb,Koer+12prb,maro+12prb} In other words, there is a well-known \textit{starting-point dependence} of the $G_0W_0$ approach,~\cite{Rink+05NJP,fuch+07prb,maro+12prb} and the best DFT starting point is usually also system dependent. It would therefore be desirable to iterate the $GW$ approach towards self-consistency to eliminate the starting point dependence. Different schemes have been developed, either
achieving self-consistency directly,~\cite{Caru+12PRB,Caruso/etal:2013_2} or by the so-called "quasi-particle
self-consistency" of Schilfgaarde et al.,~\cite{Schi+06prl} which determines the variationally best
non-interacting Green's function $G_0$. The present work will follow a simpler and
numerically more efficient approach.

We here apply the $G_0W_0$@DFT approach to quasi-linear systems of increasing length. Trans-polyacetylene  $-(C_2H_2)_n-$ is the simplest conjugated material that already exhibits the alternating set of $sp^2$-bonded carbon atoms, common to all conducting polymers, which leads to  $\pi$-delocalization of the frontier molecular orbitals dictating the behavior of the electronic gap.~\cite{Moli-Hior04POLI} We will thus use the trans-acetylene (TA) family, from small oligomers (OTAs) to the infinite polymer (TPA), as a model system to investigate the dependence of basic properties such as the IP and the HOMO-LUMO gap with localization length. To do that, we follow an  approach\cite{atal+13prb,rich+13prl} proposed recently: Building on the fact that the Kohn-Sham energy $\epsilon_N$ of the HOMO gives us the IP in exact DFT, we vary the amount of exact exchange in the Perdew-Burke-Ernzerhof hybrid functional~\cite{Perd+96JCP,Adam-Baro99JCP} (PBEh). We then pick that admixture $\alpha$ of exact exchange for which the KS-HOMO eigenvalue agrees with the quasiparticle energy from a $G_0W_0$ calculation based on the same  PBEh($\alpha$) starting point, denoted $G_0W_0$@PBEh($\alpha^{ic}$). The HOMO of PBEh is now consistent with the quasiparticle removal energy of $G_0W_0$ and for this reason we call our scheme internally-consistent \textit{ic}-PBEh. Monitoring $\alpha^{ic}$ for oligomers with increasing length then allows us to assess the length dependence of the IP, and gather information on the electronic screening.

The remainder of the paper is organized as follows: In section \ref{sec:theo}, we provide a short overview of the basic concepts of the $G_0W_0$ approximation and some technical aspects of the implementation; in section \ref{sec:res} we present our results, starting from the ground-state DFT calculations to obtain the geometrical models for the oligomers; we next present and discuss the internally-consistent model, applied to study the ionization potential of TA oligomers, with special focus on the length dependence. Finally, we draw our conclusions in section  \ref{sec:conc}.

\section{Theoretical background}
\label{sec:theo}

In many-body perturbation theory the single particle excitation energies are the solutions of the quasiparticle equation
\begin{equation}
\label{Eq:qp}
 [-\frac{\nabla^2}{2} + v_{ext}(\bfr) + v_{H}(\bfr)]\psi_{n\sigma}(\bfr) +
 \int d\bfr'\Sigma_\sigma(\bfr,\bfrp;\epsilon_{n\sigma}^{qp})\psi_{n\sigma}(\bfrp) = \epsilon_{n\sigma}^{qp}\psi_{n\sigma}(\bfr) \, ,
\end{equation}
where $v_{ext}$ corresponds to the external potential created by the nuclei, $v_{H}$ is the Hartree
potential, $n$ is a state index and $\sigma$ the associated spin. The non-local complex self-energy operator $\Sigma$ contains all electron-electron interaction effects beyond the Hartree mean field. In practice the self-energy needs to be approximated and we here adopt Hedin's $GW$ approximation~\cite{hedi65pr,bech15book} at the one-shot level
  \begin{equation}
     \label{Eq:Sigma_GW}
     \Sigma^{GW}_\sigma(\bfr,\bfrp,\epsilon) = \frac{i}{2\pi}\int d \epsilon'
           G_0^\sigma(\bfr, \bfrp, \epsilon+\epsilon')
      W_0(\bfr, \bfrp, \epsilon')e^{i\epsilon\eta}
  \end{equation}
where $\eta$ is an infinitesimal positive number.  $W_0$ is the screened Coulomb interaction
  \begin{equation}
\label{Eq:W0}
      W_0(\bfr, \bfrp, \epsilon) = \int d\bfrpp \varepsilon^{-1}(\bfr, \bfrpp, \epsilon)
           v(\bfrpp-\bfrp) \, ,
  \end{equation}
where $v(\vecr-\vecr')=1/\abs{\vecr-\vecr'}$ is the bare Coulomb interaction, and $\varepsilon^{-1}(\vecr,\vecr'';\epsilon)$ the inverse dielectric function. The latter can be written in terms of the polarizability
\begin{equation}
\label{Eq:eps_RPA}
  \varepsilon(\vecr,\vecr',\epsilon)=  \delta(\vecr-\vecr') -
                             \int \! d\vecr'' v(\vecr-\vecr'')
			     P_0(\vecr'',\vecr';\epsilon)
\end{equation}

\noindent with

\begin{equation}
\label{Eq:P}
  P_0(\vecr,\vecr';\epsilon)=-\frac{i}{2\pi} \sum_\sigma \int d\epsilon' e^{i\epsilon' \eta}
            G_0^\sigma(\vecr,\vecr';\epsilon+\epsilon')  G_0^\sigma(\vecr',\vecr;\epsilon').
\end{equation}
Finally,  $G_0$ is calculated from the eigen-energies and wavefunctions of a preceding DFT or HF calculation
\begin{equation}
\label{Eq:G0}
G_0^\sigma(\vecr,\vecr';\epsilon)=\sum_{n}\frac{\psi_{n\sigma}(\bfr) \psi_{n\sigma}^*(\bfrp)  }{\epsilon-(\epsilon_{n\sigma} + i \eta \: \mathrm{sgn}(\epsilon_F-\epsilon_{n\sigma}))}.
\end{equation}

Making the additional approximation that the quasiparticle wave functions equal the Kohn-Sham states, we can simplify Eq.  ~\ref{Eq:qp}  and write for the real part of the quasiparticle energies
\begin{equation}
\label{eq:qp_2}
   \epsilon_{n\sigma}^{qp}= \epsilon_{n\sigma}^{\rm KS}+ \Re\brokt{\psi_{n\sigma}}{\Sigma_\sigma^{GW}(\epsilon_{n\sigma}^{qp})-v_{xc}}{\psi_{n\sigma}} = \epsilon_{n\sigma}^{\rm KS} + \Delta_{n\sigma}^{qp},
\end{equation}
where $v_{xc}$ is the exchange-correlation potential of the underlying DFT (or HF) calculation and $ \Delta_{n\sigma}^{qp}=\Re\brokt{\psi_{n\sigma}}{\Sigma^{GW}(\epsilon_{n\sigma}^{qp})-v_{xc}}{\psi_{n\sigma}}$ the $G_0W_0$ or quasiparticle correction. Equations~\ref{Eq:Sigma_GW} to \ref{eq:qp_2} illustrate that $ \Delta_{n\sigma}^{qp}$ and therefore the quasiparticle energies depend on the DFT functional used in the preceding calculation.

As stated in the Introduction, in exact DFT the HOMO level of a finite system gives the IP and therefore the self-energy correction  $ \Delta_{\rm HOMO}^{qp}$ is zero (for any other level no such statement holds). In standard approximations to the exchange-correlation functional the IP is typically not given accurately, because of the self-interaction error. Minimizing  the absolute value of $ \Delta_{\rm HOMO}^{qp}$ through optimization of $\alpha$ therefore implies that the self-energy correction to the HOMO level should be as small as possible.  Alternatively, we could stay entirely within DFT and enforce the linearity of the DFT total energy with respect to the occupation of the HOMO state \cite{Perd+82PRL} to obtain $\alpha$. However, this is not the scope of this paper and we will defer a discussion of the deviation of the straight line behavior and the internally consistent $GW$ scheme to a forthcoming paper.

We emphasize that $\alpha$ is not related to a shift of the chemical potential ($\epsilon_s$) that was originally proposed by Hedin,\cite{hedi65pr} who observed that if introduced in $G_0W_0$ calculations it would model some effects of fully self-consistent GW calculations. The shift $\epsilon_s$ is also implemented in the $GW$ space-time code\cite{rieg+99cpc} and has negligible effects on the quasiparticle energies of semiconductors and insulators. This observation by Rieger et al.\cite{rieg+99cpc} is in line with the findings by Pollehn et al,\cite{poll+98jpcm} who observe differences between $G_0W_0$, shifted $G_0W_0$ and self-consistent $GW$ only in the satellite spectrum of their Hubbard clusters and not in the quasiparticle peaks.

To summarize this section, in the internally consistent $GW$ scheme we explore the space of possible $G_0$ starting points spanned by the PBEh hybrid functional, and use the $\alpha$ parameter to traverse this space. In practice, we start from the same hybrid functional\cite{Perd+96JCP} model of Perdew, Ernzerhof and Burke
\begin{equation}
\label{eq:ic_alpha}
E_{xc}= \alpha E^{\rm EX}_x + (1-\alpha)E^{\rm PBE}_x+E_c^{\rm PBE},\ \ \ 0\le\alpha\le 1\\
\end{equation}
where $E^{\rm EX}$ denotes the exact-exchange energy and $E^{\rm PBE}_x$ and $E_c^{\rm PBE}$ are the PBE exchange and correlation energy,~\cite{perd+96prl} respectively.  There the suggestion for $\alpha$  is 0.25, focusing on atomization energies of a set of molecules. Here we follow a different rationale, and thus we perform a series of PBEh calculations for different values of $\alpha$ for the same molecule, and use the Kohn-Sham eigenvalues and orbitals as input for subsequent $G_0W_0$ calculations.  We find that $| \Delta_{\rm HOMO}^{qp}|$ can be minimized by just a few single shot $G_0W_0$ calculations.

\section{Results and Discussion}
\label{sec:res}

In what follows we present first our results for TPA and OTAs obtained through different standard DFT functionals and HF, and discuss the convergence of (mean-field) electronic properties with conjugation length. Next we analyse the $G_0W_0$@DFT and $G_0W_0$@HF results for differently sized OTAs, and proceed to the discussion of the internally-consistent procedure and the effects on the electronic structure in general.

\subsection{Starting-point calculations}

Our calculations are done for oligomers ranging from \textit{n}=1 (ethylene) to \textit{n}=80 double bonds OTA(80) and for the infinite TPA chain, over a single set of geometrical structures for all adopted functionals, so that we can evaluate the effect of each functional on the electronic properties independently from the effect on the structure. All calculations are performed using the FHI-aims code,~\cite{blum+09cpc,ren+12njp} which has the advantage of including all electrons, a feature of basic relevance in our case as we will use the core-level energies explicitly for level alignment of different oligomers.  Additionally, FHI-aims offers the possibility to calculate infinite periodic as well as finite systems with the same underlying approximations\cite{havu+09jcop} (e.g., basis sets, integration grid). FHI-aims is written with numerical atomic-centered orbital basis sets, organized in so-called ``\emph{tiers}'' of basis sets, providing excellent convergence of density-functional based total energies even for complex structures, and sufficient convergence of $G_0W_0$ results.~\cite{blum+09cpc,ren+12njp} For geometric structure determination, we use the high accuracy \emph{tier} 2 basis set  (Table 1 of Ref. \onlinecite{blum+09cpc}, 39 basis functions for C and 15 for H), and the calculations were carried out using the DFT functional of Perdew, Burke and Ernzerhof~\cite{perd+96prl} (PBE) augmented by the Tkatchenko and Scheffler van der Waals scheme (vdW$^{TS}$).~\cite{Tkat-Sche09PRL} Unless otherwise stated, we select the \emph{tier} 3 basis set   (55 basis functions for C and 31 for H) to evaluate densities of states and quasiparticle energies.
In the case of $\Delta$SCF calculations we work within the spin-restricted, i.e. non-spin-polarized approximation to evaluate the total energies for the ionized systems.

We first optimize the atomic coordinates for a TPA chain employing periodic boundary conditions, with the lattice constant along the chain direction fixed at the crystalline bulk value of \textit{c}=2.457 \AA{}, as measured by X-ray scattering experiments.~\cite{Finc+82PRL} To simulate a single isolated infinite-polymer chain, the polymer backbone is placed in the \textit{(x,z)} plane (the converged ground-state geometry of TPA is planar), and the lattice parameters perpendicular to the chain direction are set to a large value (\textit{a}=\textit{b}=25 \AA{}) in order to minimize the interaction between the chains in neighbor cells. A \textit{k}-point mesh of 1x1x10 is used in the optimization procedure. For these specific settings, the resulting carbon-carbon bond distances are 1.362 \AA{} for the double bond  (C$=$C) and 1.423 \AA{} for the single bond (C$-$C), i.e., we obtain the expected dimerisation of the polymer backbone. The  C$-$H bond length is 1.095 \AA{} and the C$-$C$-$C angle is found to be 123.8$^\circ$. These structural parameters are quite similar to those found in previous theoretical studies using different functionals both within an oligomer approach~\cite{Lima+09JCP} or solid state calculations~\cite{Laci+12JCP,Hira+98PRB} and also compares favorably with experimental results.~\cite{Yann-Clar83PRL}

\begin{figure}[!ht]
 \begin{center}
    \includegraphics[scale=0.4]{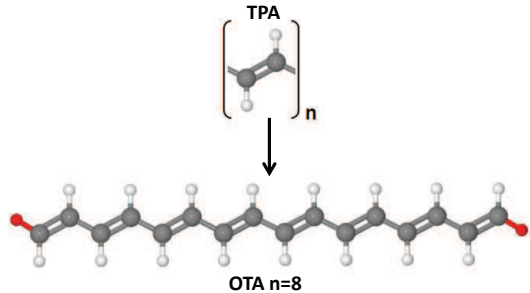}
 \end{center}
  \caption{(Color online) Schematic representation of the infinite polymer unit cell (TPA, top) and a finite model oligomer (OTA8, bottom), built by repetition of the unit cell. The hydrogen atoms added to saturate the oligomer chain and the resulting C-H distance are highlighted in red.}\label{fig_schem_tpa-to-ota8}
\end{figure}

The DFT equilibrium structure of the isolated TPA chain is then used as input to build a series of linear oligomeric chains OTA(\textit{n}). Since the focus of the present work is the length dependence of the electronic properties of oligomers, we keep the relative atomic coordinates fixed at the infinite chain result described above, and perform a further optimization only for the C$-$H distance of the end-cap CH$_2$ groups. The details of the geometry are not our main focus, as long as the geometry is consistent. We thus keep the atomic coordinates of the PBE+vdW$^{TS}$ optimization for each oligomer and apply different electronic structure approaches to these geometries. We first compare the following DFT functionals with HF: local-density approximation as parametrized by Perdew and Zunger (LDA-PZ),~\cite{perd-zung81prb,cepe-alde80prl} PBE, and PBE0~\cite{Perd+96JCP}. Then we perform  $G_0W_0$ calculations on top of these DFT functionals and on top of HF. We also apply our internally consistent scheme. The calculations are performed for the finite oligomers with up to 30 double-bonds (\textit{n}=30), which allows us to examine the length dependence of the frontier energy levels.

\begin{figure}[h]
 \begin{center}
\includegraphics[scale=0.8]{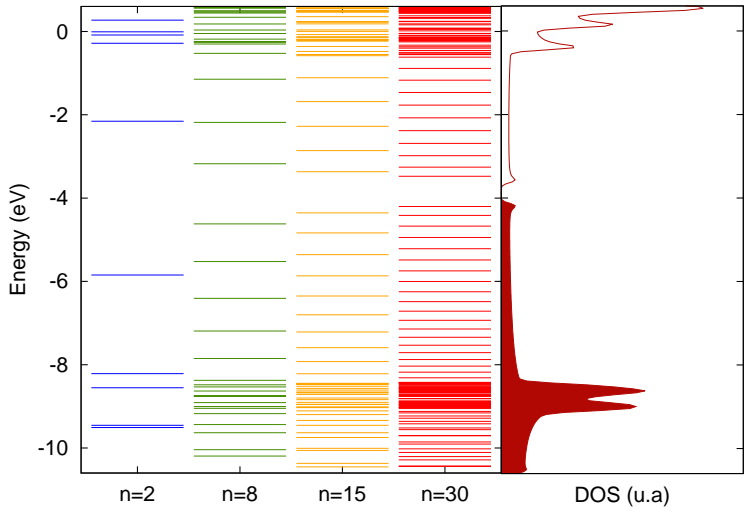}
  \caption{Discrete DFT energy level spectra (obtained with the PBE functional) of trans-acetylene oligomers (OTAs), compared to the density of states (DOS) of the 1D infinite polymer (TPA), calculated explicitly for the periodic model; here we use a gaussian broadening of 0.05 eV for the DOS. Spectra aligned at the average of the core levels ($C_{1s}$) with those of OTA(50).}\label{spectraPBE}
\end{center}
\end{figure}

Discussing first the results obtained with the standard PBE functional, we show in Fig.~\ref{spectraPBE} the KS energy-level spectra obtained for a selected series of oligomers, and the density of states (DOS) for the infinite polymer chain. To align the levels of all systems, oligomers and TPA, we use the average of the core $C_{1s}$ levels of each chain, that are then aligned at the value for the long oligomer OTA(50). As the chain length increases, we see the expected behavior of HOMO-LUMO gap closure, that converges to a small energy gap for the isolated TPA, in agreement with literature results.~\cite{rohl-loui99prl,rohr+06prb} The main features of the continuum density of states of the polymer (i.e, the width of $\pi$ HOMO and LUMO bands, and the position of localized $\pi$-states) starts to be visible for chains with $\sim$\textit{n}=15 double bonds, in agreement with previous theoretical estimates.~\cite{Hutc+03PRB}

\subsection{IP of the \emph{trans}-acetylene oligomer series}

The scaling of physical properties of finite conjugated oligomers as a function of chain length has been extensively studied and modeled in experimental and theoretical works,~\cite{Hutc+03PRB,Yang+04sm,Chi-Wegn05mmrc,Zade-Bend06OL,Gier+07AM,Peac+07JPCA,Li+12PSSB,Torr+12JPCA} aiming to predict properties of polymeric materials using different extrapolation models. We thus move now to the comparison with experimental results, and summarize in Table \ref{table_ips} the results for shorter oligomers, for which experimental data are available.~\cite{Beez+73HCA,Kimu81Handbook,Ball+84JACS,Rade+01JMS} We first list the values coming directly from the negative of the HOMO eigenvalue (columns on the left) using the aforementioned different mean-field methods (KS and HF). Next we list  the $G_0W_0$ results for the corresponding starting point, and finally the results from the internally-consistent PBEh (columns on the right). We include also specific literature results obtained with the often-employed hybrid functional B3LYP.~\cite{Salz+97JCC} The experimental values are listed in the central column.

Considering first the comparison between experiment and the LDA and PBE eigenvalues ($-\epsilon^{N}_N$) we see that, as expected, the gas-phase IP of all oligomers is strongly underestimated. The agreement with experiment is only slightly improved by the hybrid functionals (PBE0 and B3LYP), while HF values are already very close.

\begin{table}
\begin{center}
{\caption{Ionization potential of acetylene oligomers, calculated  at different theoretical levels: $-\epsilon_N$ negative of the KS (or HF) single-particle HOMO energy (left columns), our results for quasi-particle energies obtained through $G_0W_0$@DFT (right-columns), sc$GW$@HF, and from the internally-consistent procedure, see text (rightmost column). Experimental data included for comparison at the center. All energies in eV. The mean absolute error for each functional as compared to the experimental values (\textit{n}=1-4) is included in the last row.}\label{table_ips}}

\begin{tabular}{l cc cc c |c| cccc c c}
  \hline
      &  \multicolumn{2}{c}{DFT-KS} & \multicolumn{2}{c}{DFT-GKS} & HF  & Exp & \multicolumn{4}{c}{$G_0W_0$} & scGW & \textit{ic}PBEh  \\
  \textit{n}   &  LDA  &  PBE  &  PBE0  &  B3LYP~\cite{Salz+97JCC}  & & &  LDA  & PBE   &   PBE0   &   HF   &  HF  \\
  \hline
  1  &   6.85  &  6.66   &  7.77  &  7.26  & 10.10 &  10.51~\cite{Kimu81Handbook}  &  10.20  &  10.25  &  10.36  &  10.70  & 10.02 &  10.44 \\
  2  &   5.95  &  5.75   &  6.68  &  6.23  &  8.60 &   9.07~\cite{Kimu81Handbook}  &   8.65  &   8.62  &   8.83  &   9.25  & 8.47 &  8.97 \\
  3  &   5.51  &  5.31   &  6.13  &  5.69  &  7.82 &   8.29~\cite{Beez+73HCA}       &   7.78  &   7.74  &   8.00  &   8.48  & 7.65 &  8.18 \\
  4  &   5.26  &  5.04   &  5.80  &  5.36  &  7.33 &   7.79~\cite{Ball+84JACS}      &   7.20  &   7.17  &   7.47  &   8.01  & 7.15 &  7.69 \\
  5  &   5.09  &  4.87   &  5.57  &  5.14  &  7.01 &   7.00~\cite{Rade+01JMS}$^{*}$ &   6.84  &   6.80  &   7.10  &   7.69  &  &  7.36 \\
  6  &   4.97  &  4.75   &  5.41  &  4.97  &  6.78 &    ---                         &   6.56  &   6.50  &   6.81  &   7.46  &  &  7.12 \\
  8  &   4.59  &  4.81   &  5.19  &  4.75  &  6.48 &    ---                         &   6.17  &   6.09  &   6.44  &   7.16  &  &  6.79 \\
  \hline
 MAE &   3.03  &  3.23   &  2.32  &  ---  &  0.45  &  ---                           &   0.46  &   0.47  &   0.25  &   0.19  & 0.59 &  0.10 \\
  \hline
\end{tabular}
\end{center}
$^*$Experimental (gas-phase) data available only for polyenes with terminal \textit{tert}-butyl groups; in this Table, our calculated values for \textit{n}=5 are also for butyl-terminated molecules.

\end{table}

\begin{figure}[!ht]
 \begin{center}
    \includegraphics[scale=0.8]{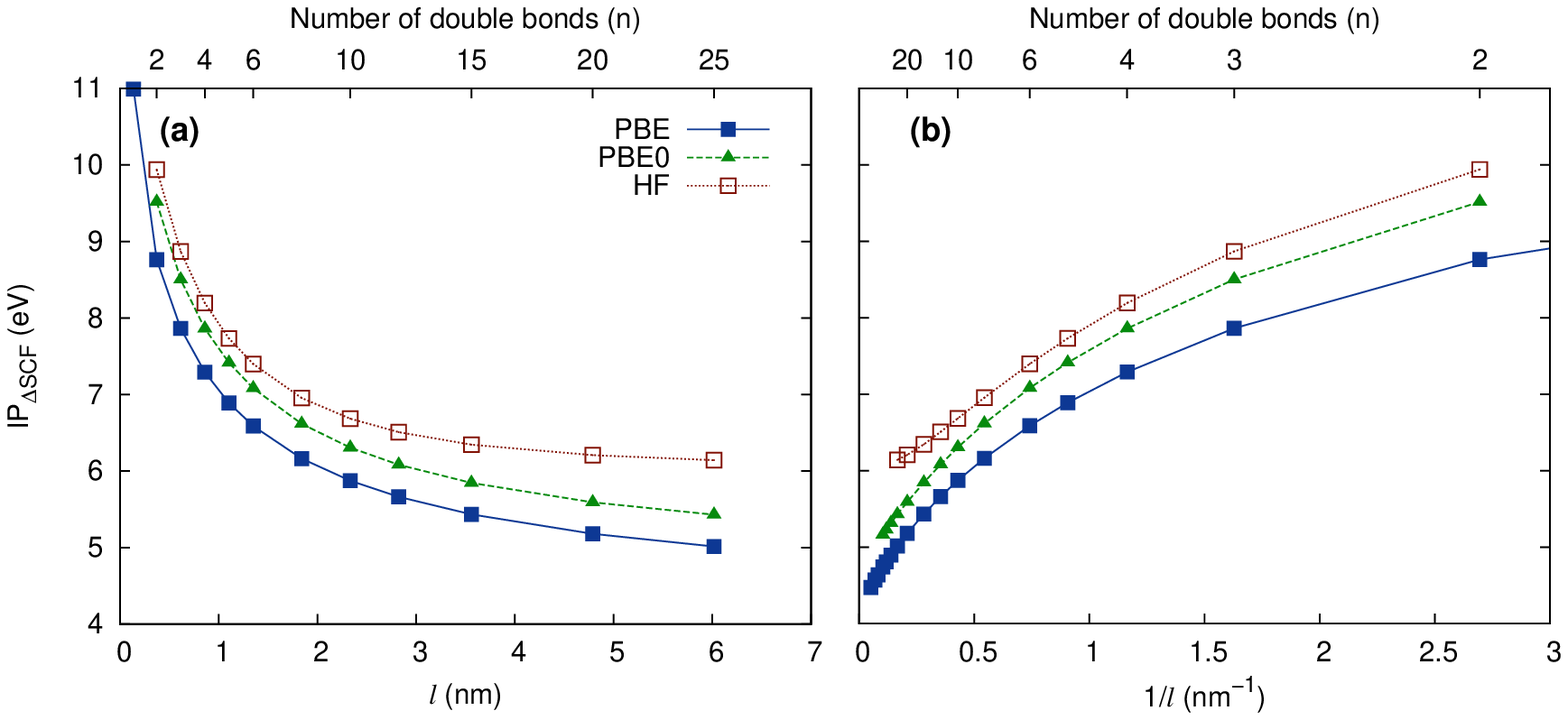}
    \includegraphics[scale=0.8]{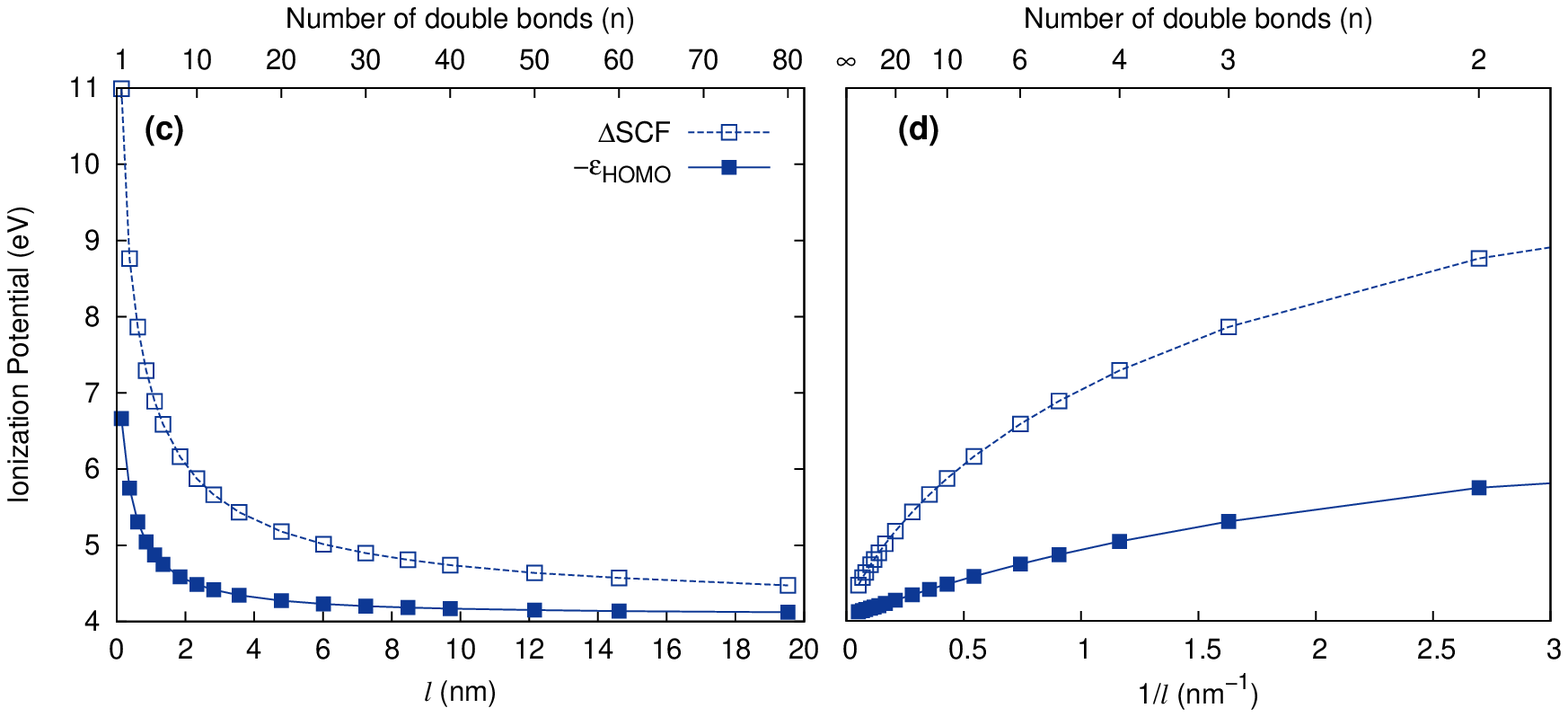}
 \end{center}
  \caption{(Color online) Top panel: Evolution of the first ionization potential of OTA series as a function of (a) chain length and (b) inverse chain length, up to \textit{n}=30 double bonds, calculated through the $\Delta$SCF approach with the different methods PBE, PBE0 and HF. Bottom panel: Comparison of the negative HOMO energy and the $\Delta$SCF approach obtained with the PBE functional, in the (c) chain length and (d) inverse chain length representation, up to \textit{n}=80 double bonds. The lines are just guides for the eye.}\label{fig_ips-dscf}
\end{figure}

In Fig.~\ref{fig_ips-dscf}(a,b) we select PBE, PBE0, and HF and now consider the evolution of the IP obtained from the difference in total energies  (IP$_{\Delta SCF}$). In Fig.~\ref{fig_ips-dscf}(c,d) we directly compare the PBE HOMO eigenvalue with IP$_{\Delta SCF }$ for the PBE functional. We first note that IP$_{\Delta SCF}$ from HF and DFT differ by a few electronvolts for small and medium sized molecules, but also that this difference tends to increase with oligomer length. The IP$_{\Delta SCF}$ calculated with PBE (or PBE0) decreases quite fast with chain length. Indeed, we can see from Fig.~\ref{fig_ips-dscf} that the slope of DFT-IP$_{\Delta SCF}$ versus $1/l$  increases with chain length, thus the value of the IP does not stabilize at longer chain lengths. A different trend is seen for HF, that is, the IP$_{\Delta SCF}$ calculated with HF  exhibits a decrease of the slope with growing oligomer length, a feature that can be seen more clearly following the inverse-length dependence. The fact that the slope in Fig.~\ref{fig_ips-dscf} increases for larger lengths in PBE and PBE0 must be attributed to the semilocal part and not to the non-local exchange part, because it does not happen for HF. The difference between the DFT and HF IP$_{\Delta SCF}$ reaches more than 1 eV in the infinite chain limit. We see also in Fig.~\ref{fig_ips-dscf}(c,d) that the actual value of the negative of the PBE HOMO eigenvalue approaches the IP$_{\Delta SCF}$  at the infinite chain length limit, but the slope of the two curves are quite different. We observe the same behavior for PBE0 (not shown here).

Let us now proceed to the quasi-particle picture: while KS-HOMO levels of organic molecules are usually too high for local or semi-local DFT functionals, many-body corrections introduced perturbatively via $G_0W_0$ calculations bring their values down, improving the description of IPs.~\cite{Rost+10PRB,Blas+11PRB} Our $G_0W_0$ results for the IPs in Table \ref{table_ips} illustrate that the differences in the quasiparticle energies are indeed significantly smaller than the differences in the HOMO energy for the original DFT or HF values. The corrected values are all in much better agreement with the measured values, with a mean absolute error smaller than 0.5 eV. However, contrary to the mean-field results, we now see an increasing deviation of the $G_0W_0$ IP from measured values with increasing oligomer length.

\begin{figure}[b]
 \begin{center}
\includegraphics[scale=0.8]{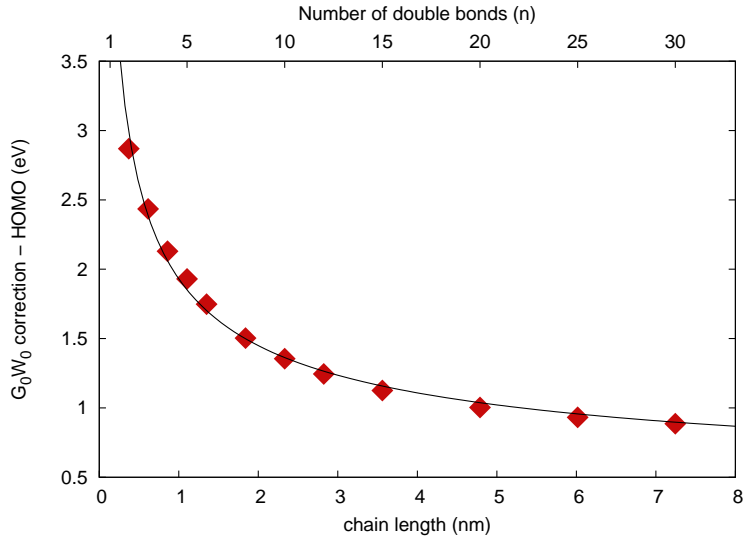}
  \caption{Length (\textit{l}) dependence of the self-energy correction on the HOMO energy $ \Delta_{\rm HOMO}^{qp}$ of the OTA chains, defined as the difference between the quasi-particle and the KS energy; results for the PBE functional. The solid line is a fit with  $ \Delta_{\rm HOMO}^{qp}=(1.6 l^{-1/2} + 0.286)$eV.}\label{fig_sec_homo}
\end{center}
\end{figure}

At this point it is illuminating to also inspect the self-energy correction $ \Delta_{\rm HOMO}^{qp}$ to the KS HOMO level, as shown in Fig.~\ref{fig_sec_homo} for PBE-based calculations. We observe that the self-energy correction decreases with chain length. This length dependence of $ \Delta_{\rm HOMO}^{qp}$ can be rationalized in terms of a length-dependent change in the screening strength of the oligomer. Given the specific $\pi$-character of the frontier orbitals, the electron density of the KS HOMO state delocalizes over the backbone of the oligomers, and when the molecular length is progressively increased from 0D ethylene towards quasi-1D oligomers, a substantial enhancement of the electronic screening is expected. The effect on the KS LUMO  is similar for these systems as we will see, and thus this is reflected in the value of the electronic gap. Niehaus et al~\cite{Nieh+05PRA} report similar conclusions for the band gap of 1D polyacenes. Also for intrinsically different systems, sp$^3$-bonded silicon nanocrystals, Delerue and co-workers~\cite{Dele+00PRL,Dele+03PRL}  observed in tight binding $GW$ calculations that the self-energy corrections to the DFT gap exhibit a smooth decreasing behavior with increasing (in that case 3D) nanocrystal size. These findings are in accordance with our results.

\subsection{Internally-consistent mixing parameter}

\begin{figure}[htb!]
 \begin{center}
  \includegraphics[scale=0.8]{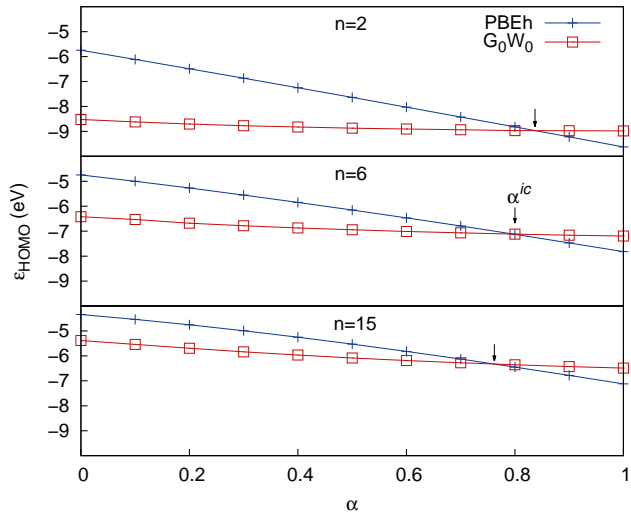}
  \caption{Evolution of the Kohn-Sham and quasiparticle HOMO energy of trans-acetylene oligomers with increase of the mixing parameter $\alpha$ of PBEh. The  parameter that satisfies the internal consistency criterion, $\alpha^{ic}$ indicated in the central panel, corresponds to the crossing-point between the curves calculated with PBEh($\alpha$) and $G_0W_0$@PBEh($\alpha$).}\label{ehomo_x_alpha}
\end{center}
\end{figure}

We now move to the choice of the mixing parameter $\alpha$ to be inserted in the PBEh functional, Eq. \ref{eq:ic_alpha}. Fig.~\ref{ehomo_x_alpha} shows the results for $G_0W_0$ quasiparticle energies compared to the original PBEh($\alpha$) KS-HOMO energies, for three chosen OTAs \textit{n}=2, 8 and 15. In this case, calculations are performed at the \emph{tier} 2 basis set level. As a consistency check we compute the $\alpha^{ic}$ value for some selected oligomers using a larger basis set, namely \emph{tier} 3, which allows for tightly converged orbital energies. The $\alpha^{ic}$ value is very stable with respect to the number of basis functions. Concerning the convergence behavior of both DFT eigenvalues and QP results we observe that the energies of the highest occupied states shift down by $\lesssim$0.1 eV when going from \emph{tier} 2 to \emph{tier} 3 basis sets. These results indicate that \emph{tier} 2 basis sets provide a good trade-off between accuracy and computational cost for the systems we study here.

Figure~\ref{ehomo_x_alpha} illustrates that the $G_0W_0$ HOMO energy depends less on the $\alpha$-parameter than the PBEh KS HOMO energy.  The intersection between the PBEh and $G_0W_0$ curves defines the internally-consistent fraction of EX namely $\alpha^{ic}$, and occurs at around $\alpha\simeq0.8$. The $\alpha^{ic}$ for 1D conjugated oligomers is thus much higher than the fraction included in most of standard hybrid functionals such as B3LYP (0.2), HSE or PBE0 (0.25).

\begin{figure}[htb!]
 \begin{center}
  \includegraphics[scale=0.8]{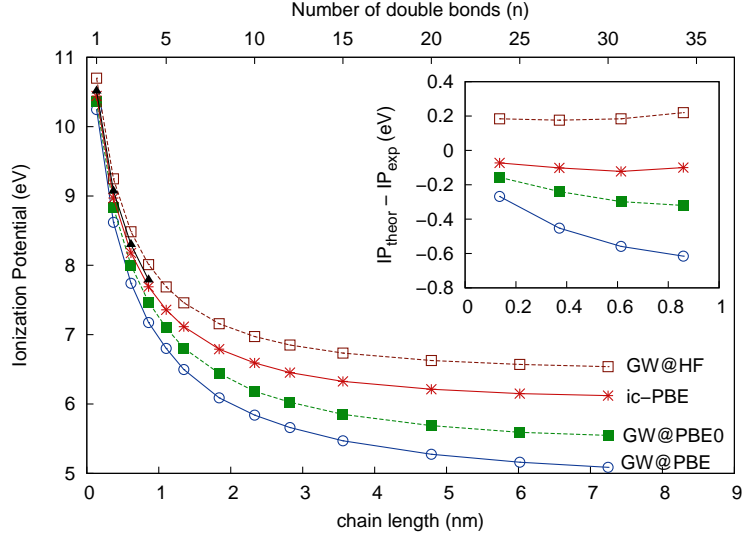}
  \caption{Chain length dependence of the first ionization potential of trans-acetylene oligomers obtained through $G_0W_0$ on different levels of mean-field methods: PBE (circles), PBE0 (solid squares) and internally-consistent PBEh (stars); Hartree-Fock (empty squares). The lines are just guides for the eye. Included are also the experimental results (solid triangles) for small oligomers.  Inset: difference between the calculated and experimental IP values for the small oligomers, same symbols as for the IP plots.}\label{fig_GW_IP_x_length}
\end{center}
\end{figure}

Our \textit{ic}-PBEh IP values are also included in Table \ref{table_ips} (\textit{ic}-PBEh) for \textit{n}=1 to 8. \emph{We see that we obtain an  improved description of the highest occupied state, yielding IPs in good agreement with gas phase reference data although no direct constraint is imposed in our scheme to fit experiment}. Interestingly, the deviation of the \textit{ic}-PBEh IPs from experiment remains approximately constant when increasing the oligomer length. This is shown in the inset of Fig.~\ref{fig_GW_IP_x_length}. We have also performed self-consistent $GW$ (sc$GW$) for reference, using the sc$GW$ implementation in FHI-aims \cite{Caru+12PRB,Caruso/etal:2013_2}. The results are included in Table \ref{table_ips}. The sc$GW$ IPs are consistently lower than experiment, $G_0W_0$@PBE0 and \textit{ic}-PBEh. This observation is consistent with recent benchmarks of sc$GW$ for molecules \cite{maro+12prb,Caru+12PRB,Caruso/etal:2013_2,caru+14prb,hell+15prb} and can be attributed to the pronounced deviation from the straight-line error (DSLE, also known as many-body self-interaction error) of sc$GW$ \cite{hell+15prb,dauth+tobe}. In $G_0W_0$, the DSLE can be reduced (or even eliminated) by an optimal starting point, which is why we here prefer to work with the \textit{ic}-PBEh scheme.

\begin{figure}[htb!]
 \begin{center}
  \includegraphics[scale=0.8]{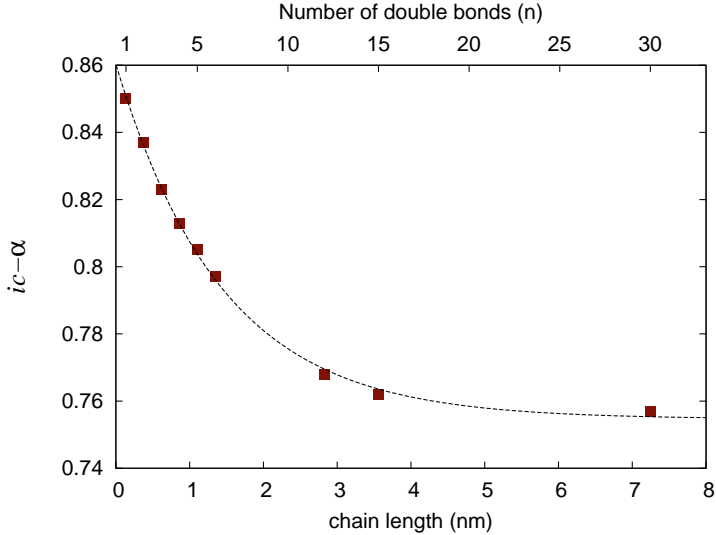}
  \caption{Size dependence of the internally consistent $\alpha$-parameter of the PBEh functional obtained for trans-acetylene oligomers. We see it decreases exponentially with chain length, as indicated by the dashed curve: the line is a fit with $\alpha^{ic}(\textit{l})=0.106e^{-0.694\textit{l}}+0.755$.}\label{fig_opta_x_length}
\end{center}
\end{figure}

We now point out that the starting-point dependence of $G_0W_0$ increases with system size, reaching $\approx$1.5 eV for $l\approx$7 nm. This tells that the $G_0W_0$ starting point is more important for longer or infinite chains. We thus plot in Fig.~\ref{fig_opta_x_length} our calculated $\alpha^{ic}$ as a function of system size. We note that it varies slightly with chain length, ranging from $\alpha\simeq 0.85$ for the ethylene molecule (\textit{n}=1) down to $\alpha\simeq 0.76$ for the longest chain with \textit{n}=30 double bonds. We fit $\alpha^{ic}(l)$ with an exponential, as indicated in Fig.~\ref{fig_opta_x_length}, which allows us to estimate the consistent fraction of exact exchange in PBEh as 75\% for the case of an isolated TPA chain (we tested also fitting with polynomials of $(1/l)$ but the errors are much larger). In contrast, K\"orzd\"orfer et al~\cite{Korz+11JCP135} recently studied long-range hybrid functionals with an additional range separation between the long-range Coulomb potential and a short-ranged effective density functional. They chose to optimize the range-separation parameter, not an overall exchange-mixing parameter $\alpha$, finding that the range-separation parameter, optimized to satisfy the DFT analog of Koopmans' theorem, strongly depends on the chain length and does not exhibit a saturation behavior for long polyene chains (\textit{n}=25) whatever the nature of the starting functional. Thus, tuning the exact exchange parameter $\alpha$ as proposed by the internally-consistent scheme~\cite{atal+13prb} is apparently a more adequate choice to predict the ionization potential for conjugated systems. As a last remark, we find that the optimal adjustment of the PBEh($\alpha^{ic}$) IP with length is obtained as IP$(l)=a+b/l+(ce^{-kl})/l$. A simpler inverse-length regression fails to reproduce the behavior at longer lengths as already discussed for energy gaps.~\cite{meie+97apol,riss+04cpl,Torr+12JPCA}

\begin{figure}[htb!]
 \begin{center}
  \includegraphics[scale=0.8]{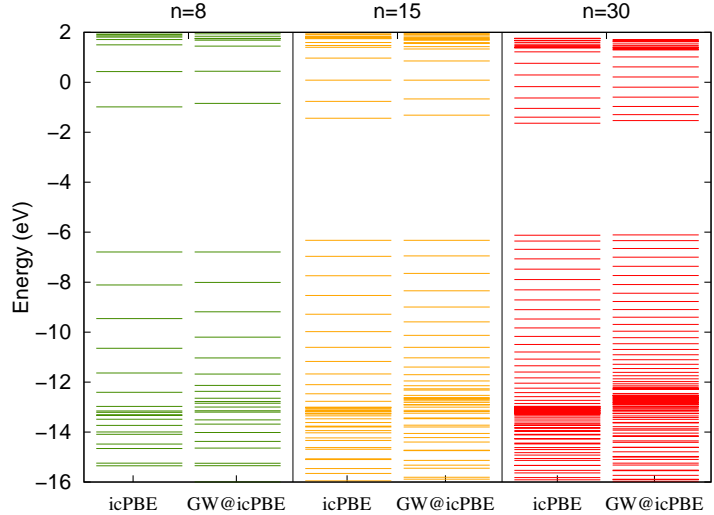}
  \caption{Comparison between the energy level spectra calculated at the PBEh($\alpha^{ic}$) and the $G_0W_0$@PBEh($\alpha^{ic}$) levels (with the internally-consistent mixing parameter) for the \textit{n}=8, \textit{n}=15 and \textit{n}=30 acetylene oligomers, close to the HOMO-LUMO gap energy window.}\label{fig_Spectra_PBEhxGW}
\end{center}
\end{figure}

Coming to more specific properties of these conjugated polymers, as pointed out before the KS LUMO shows $\pi$-conjugated symmetry and localization properties similar to the KS HOMO -- as a consequence we might expect that the behavior of the electronic gap also follows the same trend with tuning of $\alpha$. We show in Fig.~\ref{fig_Spectra_PBEhxGW} the energy spectra obtained with our consistent procedure for intermediate length oligomers. \textit{We see that the HOMO-LUMO gaps obtained through PBEh($\alpha^{ic}$) are close to the quasi-particle $G_0W_0$ values}.

\begin{figure}[htb!]
 \begin{center}
  \includegraphics[scale=0.62]{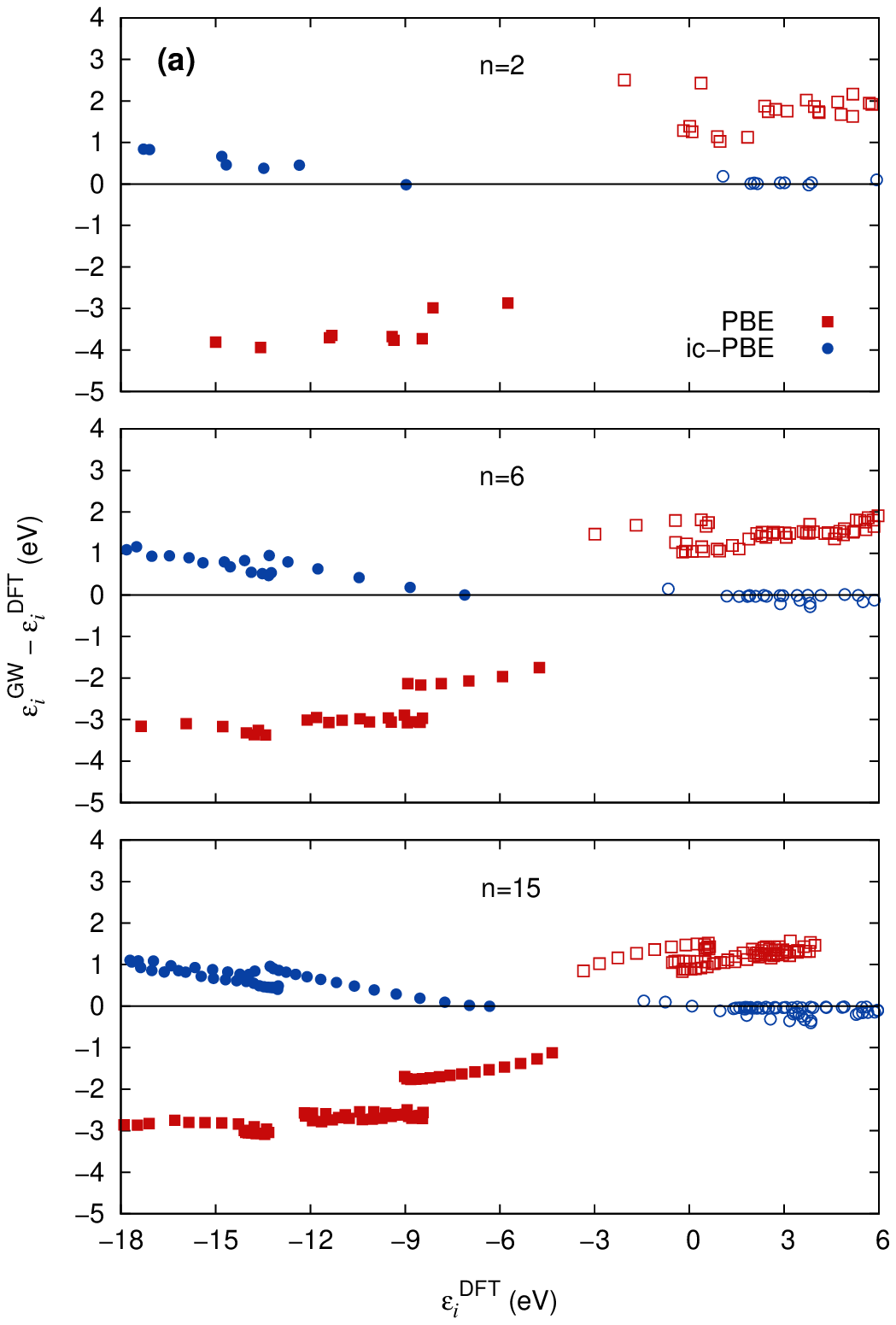}%
  \includegraphics[scale=0.62]{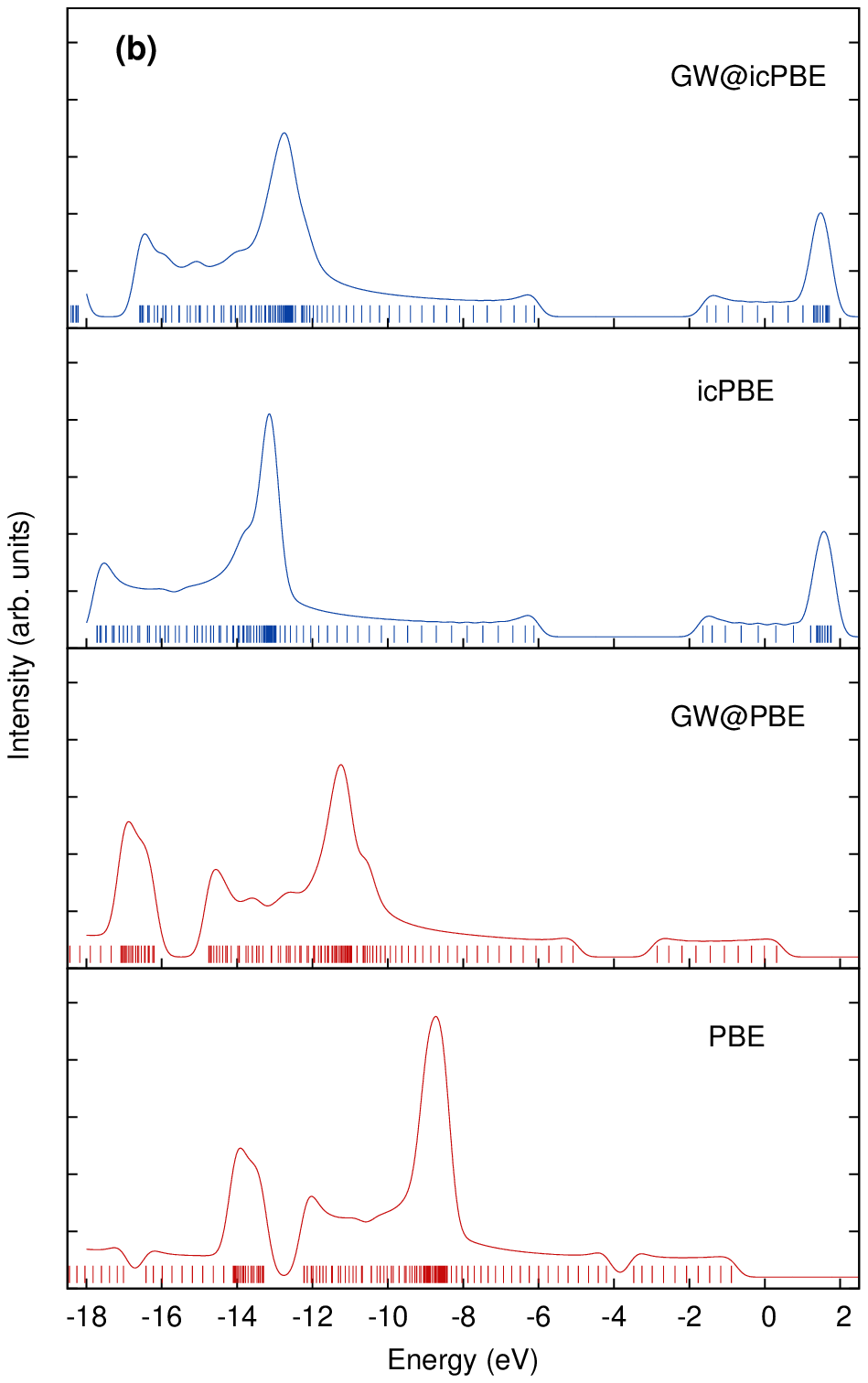}%
  \caption{Comparison of PBEh($\alpha^{ic}$) (\textit{ic}-PBE) with PBE: (a) Deviation of the Kohn-Sham (KS) eigenvalues with respect to the $G_0W_0$ quasi-particle spectrum plotted over the KS energies, for oligomer lengths \textit{n}=2, 6 and 15 double bonds. (b) Spectra obtained for the longer oligomer n=30 double bonds by the different methodologies. In (a) filled symbols indicate occupied states, empty symbols unoccupied states.}\label{fig_sic-otas}
\end{center}  
\end{figure}

We now extend our analysis to the eigenvalues for a large energy window, and not only the KS HOMO itself, and  in Fig.~\ref{fig_sic-otas} we include the results just for the PBE, \textit{i.e.} PBEh($\alpha$=0), and the PBEh($\alpha^{ic}$) starting points. In Fig.~\ref{fig_sic-otas}(a) we show the $\Delta_{\epsilon}^{qp}$ energy corrections, for the same group of oligomers in Fig.~\ref{ehomo_x_alpha}, and in Fig.~\ref{fig_sic-otas}(b) we show the complete spectra coming from the different methods, for the longer chain \textit{n}=30 double bonds. Focusing on the eigenvalues close to the frontier orbitals, we first observe that the corrections to the first unoccupied PBEh($\alpha^{ic}$) states are really negligible, even for the \textit{n}=15 double bond oligomer. Conversely, the corrections to the PBE eigenvalues are not only large ($\sim$1 eV) but also not constant as a function of energy. For the occupied states, the corrections to PBE are again large and not constant, but importantly they differ appreciably for localized and delocalized states. We recall that the peak in the density of states at $\simeq$7 eV below the KS HOMO in  Fig.~\ref{fig_Spectra_PBEhxGW} derives from localized states, thus from that energy down to the valence band minimum we start to have mixing of states with different characteristics. The self-energy correction starting from PBE, for states in this energy window, is very different from the upper window values and not predictable, while the correction to PBEh($\alpha^{ic}$) follows a smooth trend and is always less than $\sim$1 eV; this is seen more clearly in Fig.~\ref{fig_sic-otas}(b). This analysis illustrates that PBEh($\alpha^{ic}$) is a more suitable starting point for $G_0W_0$ than PBE, and thus also indicates that the PBEh($\alpha^{ic}$) spectrum is closer to experiment than PBE.

\section{Summary and Conclusions}
\label{sec:conc}

In summary, we have presented and analyzed an ``internally-consistent'' (\textit{ic}) parametrization of the PBEh($\alpha$) functional that allows us to reproduce the electronic quasiparticle energies normally obtained from $G_0W_0$ calculations for the prototypical trans-acetylene family of  conjugated systems. We show that the vertical ionization potential obtained with our optimized PBEh($\alpha^{ic}$)functional, that is, obtained through a non-empirical constraint, is always in much better agreement with available experimental values than if a simpler semilocal or standard hybrid functional (B3LYP, PBE0) is used. Furthermore, our internally consistent scheme also yields good KS LUMO energies that are consistent with $G_0W_0$, although this consistency is not a requirement in the construction of the scheme. We thus also find good agreement for electronic gaps. As a last point, we show that many-body corrections to KS MO energies close to the frontier orbitals are also smaller than those obtained for standard functionals, which allows for sound prediction of the valence photoemission spectra.
The dependence of the optimal internally consistent exchange mixing parameter $\alpha$ on the chain length is discussed and found to converge with increasing chain length. This is a significant computational advantage, as the controlled behavior of the $\alpha$ parameter should allow one to perform a single ic parametrization step for a certain class of systems and then use the corresponding PBEh($\alpha^{ic}$) functional for similar predictive simulations of the electronic structure of other, unknown systems of this class.

\begin{acknowledgments}
 The financial assistance of the Brazilian agencies CNPq and INCT-INEO are gratefully acknowledged. We also thank the action of CAPES (Brazil) and DAAD (Germany) for financial support under PhD grants. P.R. acknowledges the Academy of Finland through its Centres of Excellence Program (No. 251748 ).
\end{acknowledgments}

\bibliography{names,references_icPBE}
\end{document}